\documentclass[submission,copyright,creativecommons]{eptcs}
\providecommand{\event}{GaM 2017} 
\usepackage{breakurl}             
\usepackage{graphicx}
\usepackage{hyperref}

\providecommand{\doi}[1]{\textsc{doi}: \href{http://dx.doi.org/#1}{\nolinkurl{#1}}}

\title{Features of Agent-based Models}
\author{Reiko Heckel
\institute{Department of Informatics\\
University of Leicester, 
UK}\email{rh122@leicester.ac.uk}
\and
Alexander Kurz
\institute{Department of Informatics\\
University of Leicester, 
UK}\email{ak155@leicester.ac.uk}
\and
Edmund Chattoe-Brown 
\institute{Department of Sociology\\
University of Leicester, 
UK}
\email{\quad ecb18@leicester.ac.uk}
}
\def\titlerunning{Features of Agent-Based Models}
\def\authorrunning{R. Heckel, A. Kurz \& E. Chattoe-Brown}
\begin{document}
\maketitle

\begin{abstract}
The design of agent-based models (ABMs) is often ad-hoc when it comes to defining their scope. In order for the inclusion of features such as network structure, location, or dynamic change to be justified, their role in a model should be systematically analysed. We propose a mechanism to compare and assess the impact of such features. In particular we are using techniques from software engineering and semantics to support the development and assessment of ABMs, such as graph transformations as semantic representations for agent-based models, feature diagrams to identify ingredients under consideration, and extension relations between graph transformation systems to represent model fragments expressing features.
\end{abstract}

\section{Introduction}

The state of the art in Agent-based Modelling (ABM) is to identify phenomena of interest around which small models are developed to support specific explanations. Examples include the emergence of cooperation in competitive environments, analysed by games of Iterated Prisoner's Dilemma (IPD), or the dynamics of opinion formation in networks, using variations on the Voter model.

Often the design of such models is ad-hoc, in particular when it comes to choosing and validating their ingredient features and calibrating their parameters. For IPD we may consider features such as network structure, decision making, learning, and network change. Opinion formation models can address the impact of media and external events, network structure, network change, cognitive dissonance, psychological factors, strength of conviction, etc. A scientific approach requires both 
\begin{enumerate}
\item a systematic analysis of the choices faced when creating models, including a methodology to compare and assess their impact;
\item a way to calibrate a model's parameters and validate its results against empirical observations. 
\end{enumerate}
Without addressing these requirements, models will only be useful in answering isolated questions, making little contribution to a comprehensive understanding of the phenomena themselves, nor their relation with reality.

In this paper we consider an approach to answering the first question, the systematic analysis of the combinations of features that could be addressed in ABMs. We propose techniques from software engineering and semantics of modelling and programming languages to support the design and assessment of such features, in particular
\begin{itemize}
\item graph transformation, as semantic representation for agent-based models
\item feature diagrams, to identify ingredients under consideration and state their interdependencies
\item extensions extension relations between graph transformation systems, to represent model fragments expressing features
 \end{itemize}
 
 \section{Background}

Agent-Based Modelling~\cite{chattoe-brown2013}, hereafter ABM, is a technique for understanding social phenomena, distinct from both statistical approaches (like regression analysis) and those based on narrative accounts (like ethnography). Its distinctiveness arises from the use of a computer program explicitly representing the cognition and (inter)action of ``agents'' (simulated social actors) to explore the aggregate effects of these. This gives the usual advantages of formal approaches (when compared to the risk of incomplete, faultily reasoned or contradictory narratives) without the implausible simplifying assumptions often required of such approaches. The approach also gives rise to a distinctive methodology in which assumptions about human behaviour (based on qualitative interviews or experiments for example) can be calibrated independently of the match between real aggregate data and its simulated equivalents (validation). This approach is importantly different from the more common ``fitting'' of statistical models in which match is not achieved as a falsifiable hypothesis (as it is in ABM) but by deliberate adjustment of model parameters without a robust social interpretation. (The slope of a regression line tells us about a pattern in data. It is not clear if it tells us anything about individual behaviour.) ABM are also important because systems of interacting agents are often complex, leading to counter-intuitive outcomes in aggregate even when individual behaviours are understood. This makes casual inference from statistical regularities to individual behaviour and ``grossing up'' from individual behaviour narratives to aggregate outcomes potentially unreliable.

Despite the compelling logic of this methodology, however, large numbers of non-empirical (not calibrated or validated) ABM are still published \cite{angus2015}. These often select elements of a social phenomenon arbitrarily (for example social influence, social networks, geography and rationality) which makes them potentially non-comparable as well as non-empirical (see for example \cite{chiang2013,izquierdo2008,power} as typical examples from the huge ABM literature on the Prisoner's Dilemma). Obviously the most effective way to evaluate these non-empirical models would be using data \cite{chattoe-brown2014}. Unfortunately, for a variety of reasons (at least some of which are scientifically legitimate) this is not always feasible. Under these circumstances, new tools that allow us to systematically explore and evaluate the space of alternative models are a valuable alternative.

 \section{Example: The SIR Model}
 
\begin{figure}[h]
\centerline{\includegraphics[scale=0.3]{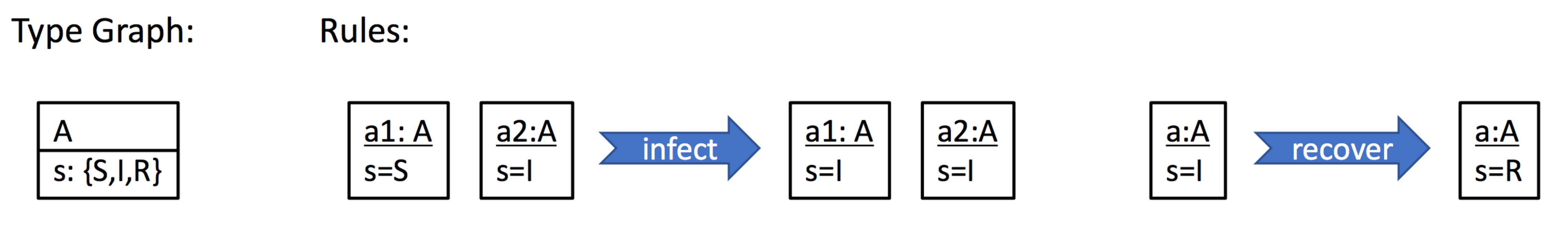}}
\caption{Basic SIR model\label{fig:SIRbasic}}
\end{figure}

An SIR model is an epidemiological model for the spread of a disease in a population of agents predicting the numbers infected with a contagious illness over time. Fig.~\ref{fig:SIRbasic} shows the most basic version of such a model. Individual agents change state from $S$usceptible to $I$nfected before becoming $R$esistant. Modelled as graph transformation system, agents and their data are specified by a type graph and their actions by graph transformation rules, as shown in Fig.~\ref{fig:SIRbasic}. The type graph defines a single node type \emph{Agent} with an attribute \emph{s} for state that can assume values $S, I$ or $R$. Rule \emph{infect} describes how an agent can change state from $S$ to $I$ in the presence of another $I$nfected agent. Rule \emph{recover} states that an infected agent can become resilient. 

\begin{figure}
\centerline{\includegraphics[scale=0.3]{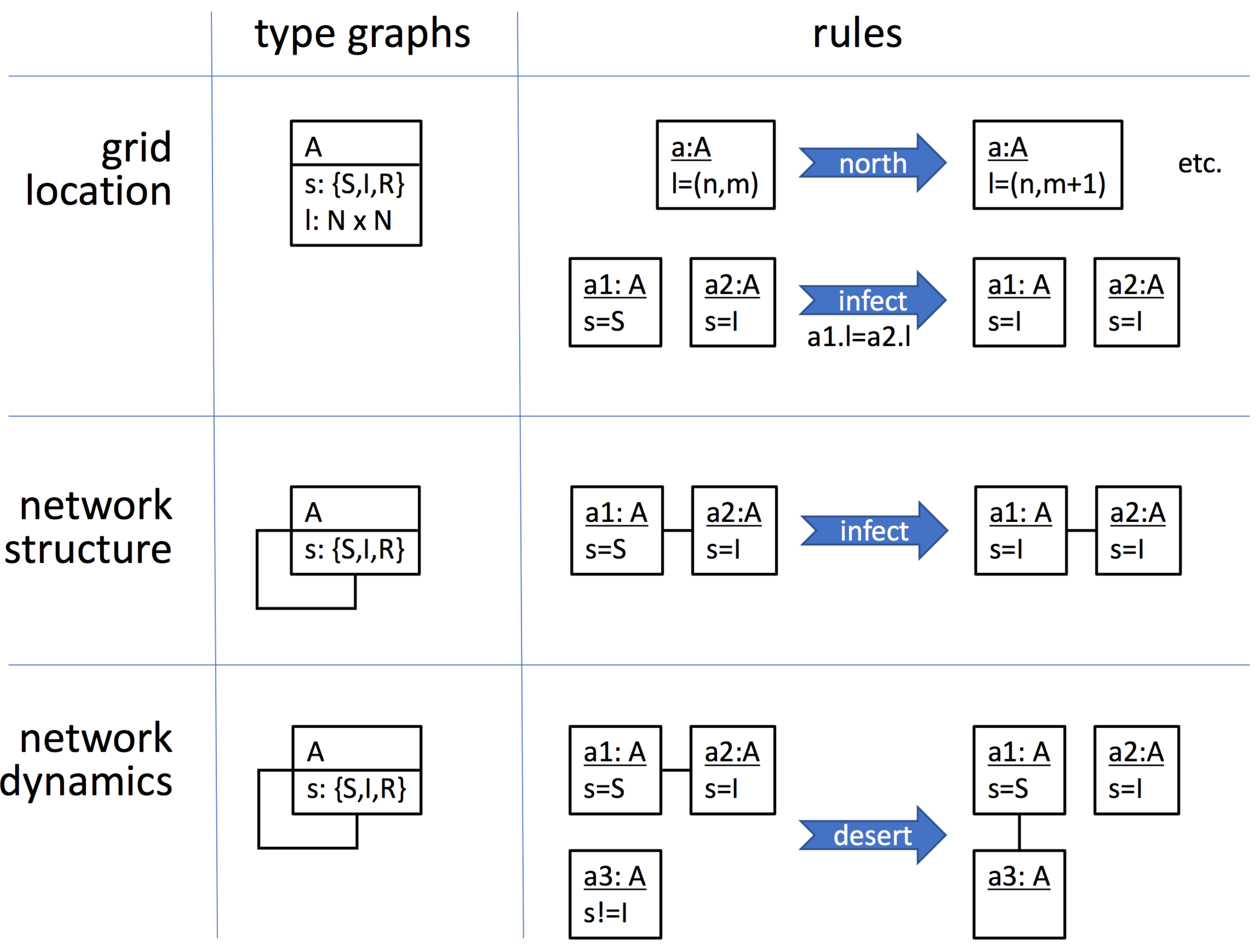}}
\caption{SIR model features\label{fig:SIRfeatures}}
\end{figure}

The basic SIR model may be able to predict levels infections in situations where agents have equal probability of interaction, but this is not in general the case. In order to get more accurate predictions we have to incorporate information about the conditions under which  infections happen. Obvious factors are location and social connections. To account for these, the model can be extended  as shown in Fig.~\ref{fig:SIRfeatures}. The \emph{location} model adds a location attribute $l$ to each agent and limits the \emph{infect} rule to such cases where both agents are in the same location. In addition, rules are added for moving in different directions. Rule \emph{north} is shown as an example. The \emph{network} model extends the base model by allowing agents to be linked. There is no movement, but the \emph{infect} rule is restricted to agents that are connected. In addition to network structure, we can introduce a \emph{dynamics} as shown in the last model. A rule is added to allow agent $a1$ to switch connection, changing their behaviour to avoid infection through $a2$.

Apart from making predictions, agent-based models represent hypotheses about the mechanisms and factors affecting social processes. For example, we can compare predictions of different versions of the model with the observed behaviour to understand which of the conditions (location, static or changeable connections) are significant to the spread of certain diseases. It is worth stressing that ABMs used for prediction or as hypotheses about real processes will be much more sophisticated than the minimal examples chosen for this paper. 

 \section{Feature Modelling and Composition}
To support the use of feature modelling in ABMs, features need to be identified, specified, composed and assessed for their relevance. We also consider the extraction of reusable features into model libraries. 

\paragraph{Feature Identification}   %
For a given phenomenon (e.g., opinion formation or disease propagation) we want to know 
\begin{itemize}
\item What are the features worth considering?
\item What configurations of features are meaningful?
 \end{itemize}
These questions are best answered studying existing literature or models, which is beyond the scope of this discussion. The result of such an analysis however can be expressed as a feature diagram $FD = (F, T)$ consisting of a set of features $F$ and a tree-like diagram $T$ over $F$ as nodes describing valid configurations $C \subseteq F$.

\paragraph{Feature Specification and Composition.}  A feature model $FM = (FD, M, m)$ is made up of a feature diagram $FD$, an underlying  model $M$ incorporating all features, usually called the 150\% model to indicate that it is not in itself a meaningful model but one requiring further restrictons, and a mapping $m$ identifying features $F$ in $M$. From this we can derive a variant $MC$ of $M$ for every valid configuration $C$.

\begin{figure}
\centerline{\includegraphics[scale=0.25]{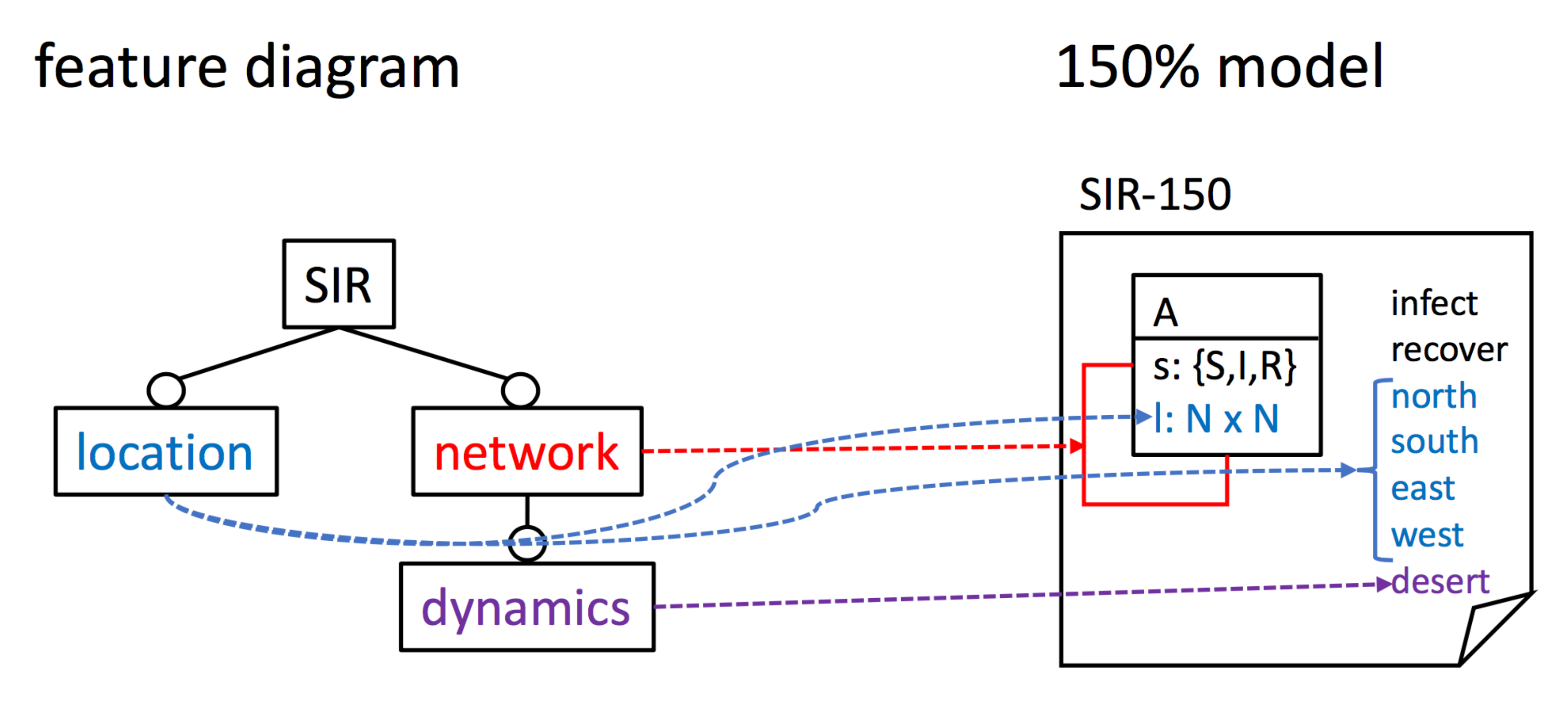}}
\caption{SIR feature model\label{fig:feature-model}}
\end{figure} 

The feature model for the SIR model and its extensions are shown in Fig~\ref{fig:feature-model}. The tree-like diagram on the left shows that the base feature \emph{SIR} can be extended by optional features \emph{location} and \emph{network} such that the latter admits a further extension by \emph{dynamics}. The mapping of features to model elements (types and rules) is illustrated in the 150\% model shown on the right. All unlabelled (black) elements belong to the base model. Elements in light blue, such as the $l$ attribute and the movement rules, belong to the \emph{location} feature, etc. 

That means, $M$ provides a combination of all possible features of the model, some of which may be mutually exclusive or redundant. There is no consideration for complex interaction of features at this stage, i.e., separate features, when added jointly, are assumed to be orthogonal. 

The mapping of features to model elements provides an interpretation of the feature tree, where nodes represent graph transformation systems and edges are extension morphisms between systems. The semantic idea is that the behaviour of the extended model can be projected onto the smaller one, i.e., extension reflects behaviour. 

Graph transformation systems can be composed along suitable morphisms. A configuration (set of features) defines a sub-tree of the feature diagram. If it exists, the colimit of this sub-diagram represents a system combining all features from this configuration~\cite{DBLP:journals/ijseke/EngelsHTE97}. 

Formally, each configuration $C$ defines a graph transformation system $GTS_C$ such that $C \subseteq D$ implies $GTS_C \subseteq GTS_D$. The inclusion is \emph{conservative} if there are no new effects on old types. That means, if rules in the extended system are reduced to the base types, their effects coincide with those of the corresponding rules in the base system. In this case, behaviours in $GTS_D$ can be projected onto behaviours of $GTS_C$, i.e., $GTS_C$ is a view of $GTS_D$. This is important because it allows us to compare two models in order to assess the relevance of their distinguishing features: If two models differing for a given feature do not show significantly different behaviour (as assessed, for example, by simulation), then the feature is not considered relevant to the behaviour of the model.

\begin{figure}
\centerline{\includegraphics[scale=0.3]{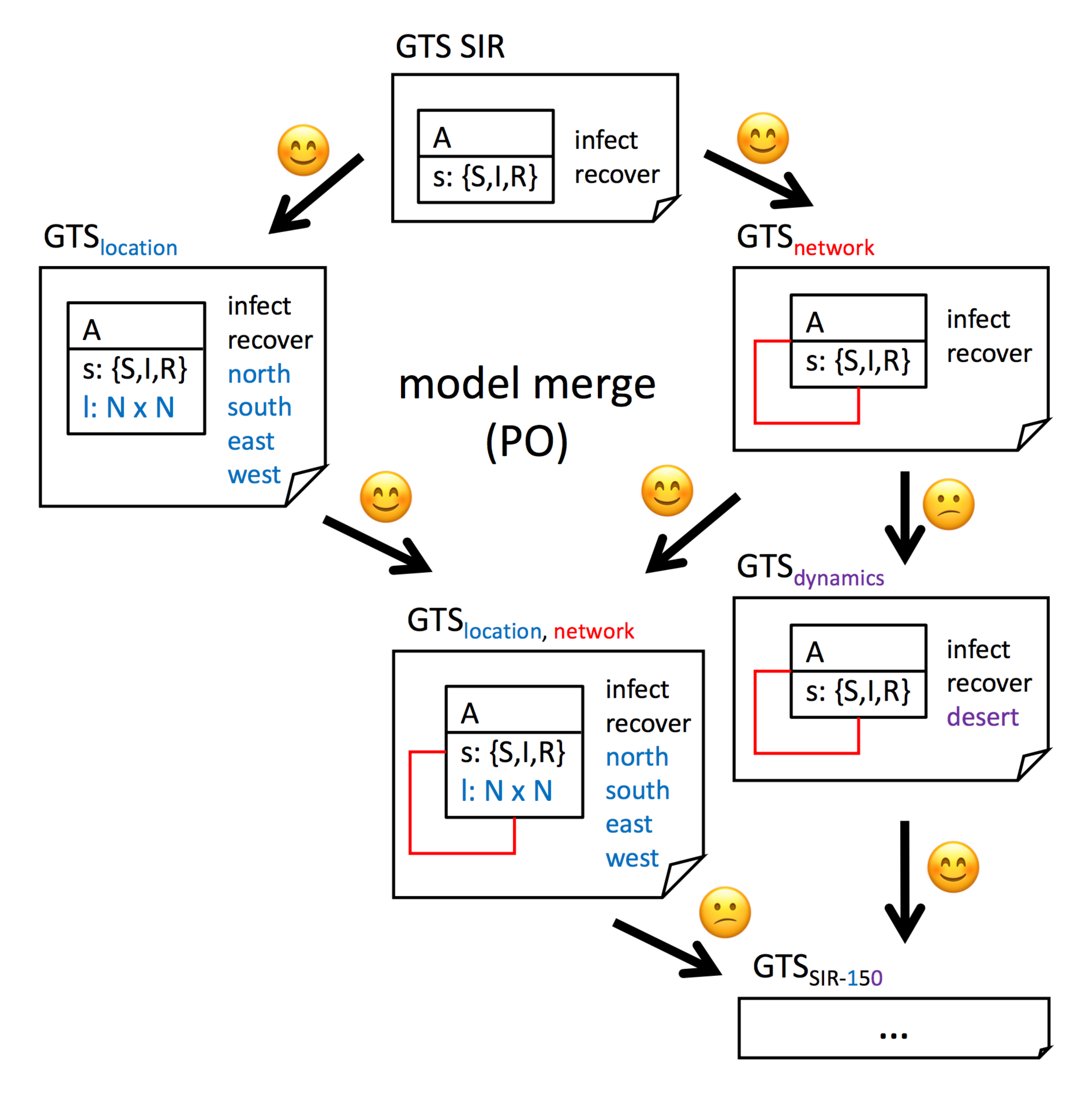}}
\caption{Variant models merging\label{fig:merge}}
\end{figure} 

Fig.~\ref{fig:merge} shows how the systems representing the different extensions of the base model are composed. Orthogonal extensions of the type graphs are merged by adding both extensions independently to the new model, while rules are simply copied so the new model inherits all  rules of both given models. It turns out that the extensions in the first merge operation are all conservative, but the extension of $GTS_{network}$ by $GTS_{dynamics}$ is not. The reason is that the \emph{desert} rule creates a new effect on instances of the existing link type, so $GTS_{dynamics}$ behaviour cannot be mapped to that of $GTS_{network}$. However, $GTS_{dynamics}$ is a conservative extension of the basic $SIR$ model. 

\section{Discussion and Future Work}

We have outlined an approach to use graph transformation and feature models to create and organise agent-based models in the social sciences.  So far we did not address the implementation of ABMs in a simulation language nor the question of calibrating a model's parameters and initial conditions, or validating its findings. 

In order to  judge if a given feature is relevant in the context of a model, we can compare model versions with or without an optional feature or containing one or another alternative feature. Based on the construction of the graph transformation systems corresponding to the relevant configurations, they are extensions of common base systems, and therefore their behaviours can be projected onto and compared from the perspective of this shared base. Combined with a suitable simulation approach this will allows us to compare results across different models and thus assess which feature configurations are relevant to understanding the phenomenon at hand.

The feature-oriented composition of systems is supported by tools such as FeatureMapper\footnote{http://featuremapper.org} which obtain a model configuration by filtering a 150 \% model. A solution specifically for graph transformation systems using the Henshin tool\footnote{https://www.eclipse.org/henshin/} is the work on variability-based graph transformations~\cite{DBLP:conf/se/0001RACTP17} supporting optional or alternative structures within rules over the same type graph. While we follow a composition-based approach where each features are mapped to separate graph transformation system related by morphisms, the variability-based solution uses annotations on model elements to map features. The difference is mainly one of notation. In both cases a composed model such as $GTS_{location, network}$ would contain one rule which propagates an infection if the agents are both at the same location and related in the network. 

Once features have been identified and assessed, we may want to reuse the components implementing them in other models. That means to abstract them from the 150\% model they were originally embedded with by specifying the assumptions by which the components can operate as parts of other models and parameterising the components to make them more widely reusable. Component concepts for graph transformation models have been studied  and may be applicable here, if they can be extended to stochastic or probabilistic models. In order to address the calibration of parameters and validation of models needed to link models to real data, notions of stochastic equivalence or similarity of models could be investigated. 

A range of techniques exist to develop, analyse and maintain feature models in software engineering. For example, reverse engineering techniques can be used to support the extraction of feature models~\cite{DBLP:conf/icse/SheLBWC11}. These are left to be explored for the usefulness in the case of agent-based modelling.



\nocite{*}
\bibliographystyle{eptcs}
\bibliography{generic}
\end{document}